\documentclass{article}
\usepackage[T1]{fontenc}

\makeatletter

\newcommand{\LyX}{L\kern-.1667em\lower.25em\hbox{Y}\kern-.125emX\spacefactor1000}

\csname @addtoreset\endcsname{equation}{section}
\newcommand{\dd}{\stackrel{\leftrightarrow}\partial}

\newcommand{\hepth}[1]{hepth/#1}
\newcommand{\npb}[3]{Nucl. Phys. B#1 (#2) #3}
\newcommand{\plb}[3]{Phys. Lett. B#1 (#2) #3}

\makeatother

\begin{document}

\begin{titlepage} 

\begin{flushright} {\small DFTT-01/00} \\ hep-th/9912284 \end{flushright} 

\vfill 

\begin{center} {

{\large On the Fixing of the \( \kappa  \) Gauge Symmetry on \( AdS \) and Flat Background: } 

the Lightcone Action for the Type \( IIb \) String on \( AdS_{5}\otimes S_{5} \).

\vskip 0.3cm

{\large {\sl }} 

\vskip 10.mm 

{\bf Igor Pesando }\footnote{e-mail: ipesando@to.infn.it, pesando@alf.nbi.dk\\

Work supported by the European Commission TMR programme ERBFMRX-CT96-004}

\vskip4mm

{\small Dipartimento di Fisica Teorica , Universit\'a di Torino, via P. Giuria 1, I-10125 Torino, Istituto Nazionale di Fisica Nucleare (INFN) - sezione di Torino, Italy} 

}

\end{center} 

\vfill

\begin{center}{ \bf ABSTRACT}\end{center} 

\begin{quote}

We explore all the possible ways of fixing the \( \kappa  \) symmetry for both branes and strings by means of a constant projector. We find that they can be classified according to their behaviour under Dirac conjugation and conjugation. This latter controls the maximum power of the fermionic variables in which the (super)vielbein can be expanded while the former determines which states cannot be described in the gauge. In particular there exists an interesting class of projectors for which vielbein are at most quadratic in the fermionic variables. As an example we compute the action for the type IIb on a \( AdS_{5}\bigotimes S_{5} \) background with a lightcone-like projector; this action reduces to the usual lightcone GS string action in the flat limit.

%\vskip 2.mm 

\vfill

\end{quote} 

\end{titlepage}

\section{Introduction.}

During the last year there has been a lot of work in finding the action for
both \( M2 \), \( D3 \) branes (\cite{MatsaevTseytlin},\cite{gruppoTo},\cite{claus})
and the type IIb superstring (\cite{MeAds5},\cite{kallosh-action},\cite{MatsaevTseytlin},\cite{MIo_AdS3},\cite{Park-Rey})
propagating on a \( AdS_{p+2}\bigotimes S_{D-p-2} \) backgrounds. All the these
actions have been derived using either the supercoset approach (\cite{MatsaevTseytlin})
with the the \( \kappa  \) symmetry fixed by the so called killing gauge (\cite{Kall-fixing})
or the supersolvable algebra approach (\cite{gruppoTo}) which automatically
yields a natural way to fix the \( \kappa  \) symmetry. It turns out that in
the cases where the actions have been worked out the two methods actually give
the same result since the killing gauge and the supersolvable gauge fix the
\( \kappa  \) symmetry in the same way.

It is purpose of this letter to explore all the possible way of fixing the \( \kappa  \)
symmerty by a constant projector of the type
\[
P=\frac{1+\Gamma }{2}\: \: \: \: \Gamma ^{2}=1\]
and discuss the advantages/disadvantages of every possible class of gauge fixings.
In order to do this we classify all the possible \( \Gamma  \)s according to
its properties under charge conjugation and under Dirac conjugation. It turns
out that the latter property is connected with the equation which describes
the string states which cannot be described while the former controls the maximum
power of the fermionic variables in the \( \kappa  \) gauge fixed expression
for the supervielbein (either \( 2 \) or \( 16 \)).

In the case of type IIb superstring propagating on a \( AdS_{5}\bigotimes S_{5} \)
we determine explicitly the equation for physical states which can be described
by the projectors which have two as the maximum power in the fermionic variables
expansion for the (super)potentials and we argue that a similar result is valid
for all the other cases.. Because of the similarity with the flat case, to which
it reduces in the infinite radius limit, we construct the action in the gauge
given by the choice \( \Gamma =\gamma _{0}\gamma _{3} \).

\section{Classifications of the constant projectors.}

As the first step we want to classify all the possible constant projectors according
to two properties which will turn out to be very important in determining the
form of the \( \kappa  \) gauge fixed fields and which physical states which
can be descrided. 

Here an in the following we assume that \( \Xi  \) are the fermionic coordinates
of the \( AdS_{p+2}\otimes S_{D-p-2} \) superspace which transform as spinor
\emph{only} under \( AdS_{p+2} \) tangent frame rotations and as a scalar for
\( S_{D-p-2} \) rotations: this is typical of the Kaluza-Klein way of decomposing
the superfields and it is different from the supercoset framework where the
fermionic coordinates transform as spinor for both \( AdS_{p+2} \) and \( S_{D-p-2} \).
Since the two approaches have to be equivalent all the results in this letter
can be in principle reformulated in the supercoset approach.

We take the projector \( P \) on the \( \Xi  \) coordinates which are to be
set to zero in order to fix the \( \kappa  \) symmetry to be written as
\begin{equation}
\label{Proj-P}
P=\frac{1+\Gamma }{2}\: \: \: \: \Gamma ^{2}=1
\end{equation}
and its complement to be \( Q=\frac{1-\Gamma }{2} \)\footnote{
In 10D the matrix \( \Gamma  \) has to satisfy another requirement: \( [\Gamma ,\Gamma _{11}]=0 \)
since the fermionic space time coordinates are chiral.
}. One could wonder why this fixes the \( \kappa  \) symmetry. A naive answer
is that it throws away half the fermionic dof as it should. A better answer
can be only given a posteriori when we can verify that the fermionic kinetic
term is not singular or when we can find an explicit \( \kappa  \) transformation
which moves a generic configuration into one which satisfies this gauge.

For a generic projector \( P \) we define the following shorthand notations:
\begin{eqnarray*}
\overline{P} & = & \gamma _{0}P^{\dagger }\gamma _{0}\\
P_{c} & = & C(\overline{P})^{T}C^{-1}
\end{eqnarray*}
where \( C \), the charge conjugation matrix, and the \( \gamma  \)s are the
proper matrices for the given spacetime, i.e. they are the \( AdS \) \( \gamma  \)
and \( C \) matrices. We define similar shorthand for the \( Q \) projector.

We can now introduce the classification according to the behaviour under the
Dirac conjugation:

\begin{eqnarray}
Type\: L: & \: \: \:  & \overline{P}=P\: \: \: \overline{Q}=Q\label{L} \\
Type\: S: & \: \: \:  & \overline{P}=Q\: \: \: \overline{Q}=P\label{S} 
\end{eqnarray}
As we show later in section (\ref{SL-section}) this classification has to do
with whether or not the left invariant forms obtained from a specific coset
elements rapresentation are quadratic in the fermionic variables (S) or up to
the sixteenth power (L).

We also divide the projectors according to their behaviour under charge conjugation:
\begin{eqnarray}
Type\: I: & \: \: \:  & P_{c}=P\: \: \: Q_{c}=Q\label{I} \\
Type\: II: & \: \: \:  & P_{c}=Q\: \: \: Q_{c}=P\label{II} 
\end{eqnarray}
This subdivision, as we will see in section (\ref{k-states-section}), has to
do with which physical states can be described by the gauge enforced by the
projector \( P \). This will be explicitly proved in the case of the string
propagating on \( AdS_{5}\bigotimes S_{5} \) but we believe it is true for
all the cases.

Using the conventions of app. \ref{app_conv} we can now make a series of tables
with all the possible \( \Gamma  \) defining the projectors up to Lorentz rotations
in the world volume of the \( Dp \) brane or in its transverse space when we
think to \( AdS_{p+2}\otimes S_{D-p-2} \) as the near horizon geometry:

\begin{itemize}
\item \( AdS_{3}\otimes S_{3} \) case:

\vspace{0.3cm}
{\centering \begin{tabular}{|c|c|c|}
\hline 
&
I&
II\\
\hline 
\hline 
S&
\( i\gamma _{2}\: ,\: i\gamma _{1} \)&
\\
\hline 
L&
&
\( \gamma _{0} \)\\
\hline 
\end{tabular}\par}
\vspace{0.3cm}

where the case \( \Gamma =i\gamma _{2} \) corresponds to the supersolvable
algebra;

\item \( AdS_{5}\bigotimes S_{5} \) case:

{\centering \begin{tabular}{|c|c|c|}
\hline 
&
I&
II\\
\hline 
\hline 
S&
\( \gamma _{0}\gamma _{3}\: ,\: \gamma _{0}\gamma _{4} \)&
\( i\gamma _{3}\: ,\: i\gamma _{4} \)\\
\hline 
L&
\( \gamma _{0} \)&
\( i\gamma _{2}\gamma _{3}\: ,\: i\gamma _{3}\gamma _{4} \)\\
\hline 
\end{tabular}\par}

We notice that the choice \( \Gamma =i\gamma _{4} \) corresponds to the supersolvable
gauge;

\item Flat 10D case:
\end{itemize}
{\centering \begin{tabular}{|c|c|c|}
\hline 
&
I&
II\\
\hline 
\hline 
S&
\( \Gamma _{0}\Gamma _{9} \)&
\( i\Gamma _{0}\Gamma _{1}\Gamma _{2}\Gamma _{3} \)\\
\hline 
L&
\( \Gamma _{6}\Gamma _{7}\Gamma _{8}\Gamma _{9} \)&
\( i\Gamma _{8}\Gamma _{9} \)\\
\hline 
\end{tabular}\par}

Similar considerations can be easily done for branes and membranes.

\section{Short and Long left invariant forms expansion\label{SL-section}.}

We now want to discuss explicitly the meaning of the \( S \) and \( L \) projectors.
To this purpose we write down an explicit representation for the coset elements
from which a nice expressions for the fermionic dependency of left invariant
forms can be derived. In particular we take as a representative for coset elements
the following expression (\cite{Tonin})
\begin{equation}
\label{L-coset_rep}
L(x,\Xi )=L_{F}(\Xi )L_{B}(x)
\end{equation}
where
\begin{equation}
\label{ferm-coset-repr-impl-S}
L_{F}=\exp \left( \begin{array}{cc}
 & (P\Xi )_{\alpha N}\\
(\overline{P\Xi })^{\beta M} & 
\end{array}\right) \exp \left( \begin{array}{cc}
 & (Q\Xi )_{\alpha N}\\
(\overline{Q\Xi })^{\beta M} & 
\end{array}\right) 
\end{equation}
 where \( \alpha ,\beta  \) are \( AdS \) spinorial indeces which label the
rows of the upper right block and \( N,M \) are column indeces on which the
\( R \) symmetry acts. 

We notice that even before fixing the \( \kappa  \) symmetry there is a big
difference between the projectors S and L, in fact type S projectors yield an
expression for \( L_{F} \) quartic in \( \Xi  \) when explicitly evaluated:
\begin{equation}
\label{ferm-coset-repr-expl-S}
G_{F}=\left( \begin{array}{cc}
\left( 1+\frac{1}{2}P\Xi ^{R}\overline{\Xi }_{R}Q\right) _{\alpha }^{.\beta } & (P\Xi )_{\alpha N}\\
(\overline{\Xi }Q)^{\beta M} & 1
\end{array}\right) \left( \begin{array}{cc}
\left( 1+\frac{1}{2}Q\Xi ^{R}\overline{\Xi }_{R}P\right) _{\alpha }^{.\beta } & (Q\Xi )_{\alpha N}\\
(\overline{\Xi }P)^{\beta M} & 1
\end{array}\right) 
\end{equation}
The same kind of computation gives a very different result for the type L projectors
since the equivalent of eq. (\ref{ferm-coset-repr-impl-S}) would read 
\begin{equation}
\label{ferm-coset-repr-impl-L}
G_{F}=\exp \left( \begin{array}{cc}
 & (P\Xi )_{\alpha N}\\
(\overline{\Xi }P)^{\beta M} & 
\end{array}\right) \exp \left( \begin{array}{cc}
 & (Q\Xi )_{\alpha N}\\
(\overline{\Xi }Q)^{\beta M} & 
\end{array}\right) 
\end{equation}
which is a product of two \( O(\Xi ^{16}) \) (in the case of maximal supersymmetry)
matrices. Because of this complexity we will not consider anymore the projectors
of type L.

These properties imply directly the maximum power of the fermionic coordinates
in the supervielbeins expansion through the formula \( L^{-1}dL=supervielbein+spin-connection \);
in particular it is immediate to verify using the \( \kappa  \) gauge fixed
vesion of eq. (\ref{ferm-coset-repr-expl-S}), i.e. setting \( P\Xi =0 \),
that the \( \kappa  \) gauge fixed supervielbeins obtained using S projectors
are \emph{at most quadratic} in the fermionic coordinates.

\section{The equation for the states which can be described by the \protect\( \kappa \protect \)
gauge fixed action\label{k-states-section}.}

In this section we concentrate on the consequences of choosing either a type
I or a type II projector in the specific case of type IIB superstring on \( AdS_{5}\bigotimes S_{5} \).
The same conclusions can essentially be reached in the flat superstring case
because the gravitino transformation rule under \( \kappa  \) symmetry is essentially
eq. (\ref{kappa-symmetry}) without the second rhs term which corresponds to
the internal \( S_{5} \) contribution. Moreover we believe that similar conclusions
can be drawn for the other cases because while the gravitino transformation
rules have additional terms in the \( \kappa  \) symmetry expression (\ref{kappa-symmetry})
the essential term is given by the first term in rhs which is common to all
the cases.

We start writing the first order action for the GS superstring propagating on
\( AdS_{5}\bigotimes S_{5} \) in order to fix the notations which we keep almost
as in (\cite{MeAds5}): 
\begin{eqnarray}
S & = & \frac{T}{2}\int _{\Sigma }\Pi ^{\widehat{a}}_{\underline{\alpha }}\: i_{*}V^{\widehat{b}}\wedge e^{\underline{\beta }}\: \eta _{\widehat{a}\widehat{b}}\: \epsilon _{\underline{\alpha }\underline{\beta }}-\frac{1}{4}\Pi ^{\widehat{a}}_{\underline{\alpha }}\Pi ^{\widehat{b}}_{\underline{\beta }}\: \eta _{\widehat{a}\widehat{b}}\: \eta ^{\underline{\alpha }\underline{\beta }}\: \epsilon _{\underline{\gamma }\underline{\delta }}\: e^{\underline{\gamma }}\wedge e^{\underline{\delta }}\nonumber \\
 &  & +\frac{1}{4}e^{i\phi }i_{*}\left( V_{+}*A\right) -\frac{1}{4}e^{-i\phi }i_{*}\left( V_{-}*A\right) \label{S-primo-ordine} 
\end{eqnarray}
where \( e^{\underline{\alpha }} \) (\( \underline{\alpha },\underline{\beta },\ldots \in \left\{ 0,1\right\}  \)
worldsheet flat indeces) are the worldsheet zweibein, \( \Pi ^{\hat{a}}_{\underline{\alpha }} \)
(\( \hat{a},\hat{b},\dots \in \{0,\ldots 9\} \) flat spacetime indeces) are
auxiliary 0-forms related by the equations of motion to \( V^{\hat{a}} \),
the ten dimensional zehnbein, by \( i_{*}V^{\hat{a}}=\Pi ^{\hat{a}}_{\underline{\alpha }}e^{\underline{\alpha }} \)
where \( i_{*} \) is the pullback on the string worldsheet \( \Sigma  \) due
to the superimmersion \( i:\: \Sigma \: \rightarrow \: AdS_{5}\bigotimes S_{5} \).
Morevover we have \( V_{\pm }*A=\epsilon _{\alpha \beta }V_{\pm }^{\alpha }A^{\beta } \)
(\( \alpha ,\beta \in \{0,1\} \) \( SU(1,1) \) fundamental irrep indeces)
with \( V_{\pm }^{\alpha } \) the \( SU(1,1) \) scalars and \( A^{\alpha } \)
the complex 2-forms; finally \( \phi  \) is an arbitraty angle we can reabsorb
redefining the complex fields \( A^{\alpha } \). 

The zweibein equation of motion gives the Virasoro constraints:
\begin{equation}
\label{Virasoro-const}
\frac{\delta S}{\delta e}=0\Longrightarrow \Pi _{\underline{\alpha }}\bullet \Pi _{\underline{\beta }}=\frac{1}{2}\eta _{\underline{\alpha }\underline{\beta }}\Pi ^{2}
\end{equation}

This action has a \( \kappa  \) symmetry which is generated by the action of
the Lie derivative with respect to a vector field \( \overrightarrow{\epsilon } \)
which is defined by the properties
\begin{eqnarray}
i_{\overrightarrow{\epsilon }}\: \psi _{N} & = & \epsilon _{N}=\Pi ^{a}_{\underline{\alpha }}\: \gamma _{a}\kappa ^{\underline{\alpha }}_{N}-i\Pi ^{i}_{\underline{\alpha }}\: J_{i}|^{P}_{N}\: \kappa ^{\underline{\alpha }}_{P}\label{kappa-symmetry} \\
\kappa _{c\underline{\alpha }}^{N} & = & -ie^{i\phi }\: L^{NP}\: \epsilon _{\underline{\alpha }\underline{\beta }}\: \kappa _{P\underline{\beta }}\label{vincolo-su-kappa} 
\end{eqnarray}
with \( J_{i}|^{P}_{N} \) and \( L^{NP} \) some functions on the 5-sphere
\( S_{5} \) (they are properly defined in app.(\ref{AdS5_conv}) where some
of their useful properties are given). 

We want now understand the relation between this vector field \( \overrightarrow{\epsilon } \)
and the associated variations \( \delta _{\kappa }\Theta _{N} \), \( \delta _{\kappa }\Theta ^{N}_{c} \)
and \( \delta _{\kappa }x \). From eq. (\ref{kappa-symmetry}) we see clearly
that the vector field \( \overrightarrow{\epsilon } \) can be written as 
\begin{equation}
\label{epslino-anholomic}
\overrightarrow{\epsilon }=\epsilon _{\alpha N}\overrightarrow{D}^{\alpha N}\; \; \; \; \psi _{\beta M}\left( \overrightarrow{D}^{\alpha N}\right) =\delta ^{\alpha }_{\beta }\delta ^{N}_{M}
\end{equation}
and if we could expand \( \overrightarrow{D}^{\alpha N} \)on the usual tangent
basis \( \partial ^{N}=\frac{\partial }{\partial \Theta _{N}} \), \( \partial _{N}=\frac{\partial }{\partial \Theta _{c}^{N}} \)
and \( \partial _{x}=\frac{\partial }{\partial x} \) then we could read the
desired variations \( \delta _{\kappa }\Theta _{N} \), \( \delta _{\kappa }\Theta ^{N}_{c} \)
and \( \delta _{\kappa }x \) since
\begin{equation}
\label{epsilon-holomic}
\overrightarrow{\epsilon }=\delta _{\kappa }\Theta _{N}\: \overrightarrow{\partial }^{N}+\delta _{\kappa }\Theta _{c}^{N}\: \overrightarrow{\partial }_{N}+\delta _{\kappa }x\: \overrightarrow{\partial }_{x}
\end{equation}

It is not very difficult to convinve oneself that starting from a coset representative
as in eq. (\ref{L-coset_rep}) we can write
\begin{eqnarray}
\delta _{\kappa }P\Theta _{\alpha N} & =M{}^{\beta }_{\alpha }\: P\epsilon _{\beta N}+M{}^{\beta }_{\alpha }\: Q\epsilon _{\beta N}+N{}^{\beta }_{\alpha }\: Q\epsilon _{c\beta }^{N}+\left[ O(\Xi ^{2})\right] _{\alpha N}\label{theta-bos} 
\end{eqnarray}
where \( M \) is a bosonic invertible matrix. The important point is that now
we can set \( Q\epsilon  \) to zero and still have an invertible bosonic matrix
connecting \( P\epsilon _{N} \)and \( \delta _{\kappa }P\Theta _{N} \): the
fermionic correction does not matter since the bosonic part is invertible and
hence it can be treated perturbatively.

In order to discuss the restrictions of the two projector types on the physical
states which can be described we start noticing that using eq. (\ref{theta-bos})
we can trade \( P\epsilon _{N} \) for \( \delta _{\kappa }P\Theta _{N} \),
i.e. we can check when we can find the \( \kappa  \) parameters (\ref{kappa-symmetry})
which allow to choose the gauge \( P\epsilon _{N}=0 \) and then we can translate
immediately the result to our actual \( \kappa  \) variation \( \delta _{\kappa }P\Theta _{N} \)
because \( P\epsilon _{N} \) and \( \delta _{\kappa }P\Theta _{N} \) are connected
by a non singular linear transformation. Obviouslly there will be some fermionic
corrections but these can always be handled because the bosonic part is invertible.
Hence the configurations where we cannot reach the gauge \( P\epsilon _{N}=0 \)
are the same where we cannot reach the true gauge \( P\Theta _{N}=0 \). Because
of this we can work with \( \epsilon ^{(+)}\equiv P\epsilon  \) and then we
can reread the results for \( P\Theta  \). 

As a first step we use eq. (\ref{vincolo-su-kappa}) into eq. (\ref{kappa-symmetry})
to eliminate \( \kappa _{N1} \) then we write down the relation between \( \epsilon ^{(+)} \)
, its charge conjugate and the \( \kappa  \) parameters for type I projectors
 
\begin{eqnarray}
\epsilon _{N}^{(+)} & = & {\cal A}_{0}|_{N}^{M}\: \lambda _{M}+ie^{-i\phi }{\cal A}_{1}|_{N}^{M}\: L_{MP\: }\: \lambda _{c}^{P}+\nonumber \\
 &  & +{\cal C}_{0}|_{N}^{M}\: \mu _{M}+ie^{-i\phi }{\cal C}_{1}|_{N}^{M}\: L_{MP}\: \mu _{c}^{P}\label{epsilon-I} \\
ie^{-i\phi }L_{NP}\: \epsilon _{c}^{P(+)} & = & {\cal A}_{1}|_{N}^{M}\: \lambda _{M}+ie^{-i\phi }{\cal A}_{0}|_{N}^{M}\: L_{MP}\: \lambda _{c}^{P}\nonumber \\
 &  & +{\cal C}_{1}|_{N}^{M}\: \mu _{M}+ie^{-i\phi }{\cal C}_{0}|_{N}^{M}\: L_{MP}\: \mu _{c}^{P}\label{L-epsilon-c-I} 
\end{eqnarray}
and for type II projectors 
\begin{eqnarray}
\epsilon _{N}^{(+)} & = & {\cal A}_{0}|_{N}^{M}\: \lambda _{M}+ie^{-i\phi }{\cal C}_{1}|_{N}^{M}\: L_{MP\: }\: \lambda _{c}^{P}+\nonumber \\
 &  & +{\cal C}_{0}|_{N}^{M}\: \mu _{M}+ie^{-i\phi }{\cal A}_{1}|_{N}^{M}\: L_{MP}\: \mu _{c}^{P}\label{epsilon-II} \\
ie^{-i\phi }L_{NP}\: \epsilon _{c}^{P(-)} & = & {\cal C}_{1}|_{N}^{M}\: \lambda _{M}+ie^{-i\phi }{\cal A}_{0}|_{N}^{M}\: L_{MP}\: \lambda _{c}^{P}\nonumber \\
 &  & +{\cal A}_{1}|_{N}^{M}\: \mu _{M}+ie^{-i\phi }{\cal C}_{0}|_{N}^{M}\: L_{MP}\: \mu _{c}^{P}\label{L-epsilon-c-II} 
\end{eqnarray}
where we have defined \( \mu _{M}=P\kappa _{M0} \) , \( \lambda _{M}=Q\kappa _{M0} \)
and \( {\cal A} \), \( {\cal C} \) by means of the relations
\begin{eqnarray*}
Q{\cal A}={\cal A}P,\: \: P{\cal A}={\cal A}Q\Longrightarrow {\cal A}_{\underline{\alpha }}|_{N}^{M} & = & \Pi ^{r}_{\underline{\alpha }}\: \gamma _{r}\: \delta ^{M}_{N}\\
Q{\cal C}={\cal C}Q,\: \: P{\cal C}={\cal C}P\Longrightarrow {\cal C}_{\underline{\alpha }}|_{N}^{M} & = & \Pi ^{u}_{\underline{\alpha }}\: \gamma _{u}\: \delta ^{M}_{N}-i\: \Pi ^{i}_{\underline{\alpha }}\: J^{i}|_{N}^{M}
\end{eqnarray*}
where \( r \) runs on a proper subset \( R \) of the spacetime index set \( \left\{ 0,1,2,3,4\right\}  \)
and \( u \) on its complement \( \left\{ 0,1,2,3,4\right\} \setminus R \);
both sets depend on the the explicit form chosen for the matrix \( \Gamma  \)
entering the definition of the projector (\ref{Proj-P}), for example
\[
\Gamma =\gamma _{0}\gamma _{3}\Longrightarrow R=\left\{ 0,3\right\} \]
We can now make the following linearly independent combinations for type I projectors:
\begin{eqnarray}
\epsilon _{N}^{(+)}\pm ie^{-i\phi }L_{NP}\: \epsilon _{c}^{P(+)} & = & {\cal A}_{\pm }|_{N}^{M}\: \left[ \lambda _{M}\pm ie^{-i\phi }L_{MP\: }\: \lambda _{c}^{P}\right] +\nonumber \\
 &  & +{\cal C}_{\pm }|_{N}^{M}\: \left[ \mu _{M}\pm ie^{-i\phi }L_{MP\: }\: \mu _{c}^{P}\right] \label{lin-comb-eps-I} 
\end{eqnarray}
 and for type II projectors:
\begin{eqnarray}
\epsilon _{N}^{(+)}\pm ie^{-i\phi }L_{NP}\: \epsilon _{c}^{P(-)} & = & \left( {\cal A}_{0}\pm {\cal C}_{1}\right) |_{N}^{M}\: \left[ \lambda _{M}\pm ie^{-i\phi }L_{MP\: }\: \lambda _{c}^{P}\right] +\nonumber \\
 &  & \pm \left( {\cal A}_{1}\pm {\cal C}_{0}\right) |_{N}^{M}\: \left[ \mu _{M}\pm ie^{-i\phi }L_{MP\: }\: \mu _{c}^{P}\right] \label{lin-comb-eps-II} 
\end{eqnarray}

Let us now discuss type I projectors only for brevity. We notice that the two
equations in eq. (\ref{lin-comb-eps-I}) are linearly independent also with
respect to the charge conjugation operation under which each equation transforms
in itsself, hence we can solve each of them independently of the other. As a
first step we could try to solve the equations (\ref{lin-comb-eps-I}) for either
\( \left[ \lambda _{M}\pm ie^{-i\phi }L_{MP}\lambda _{c}^{P}\right]  \) or
\( \left[ \mu _{M}\pm ie^{-i\phi }L_{MP}\mu _{c}^{P}\right]  \). We find that
in both cases the necessary condition for the non singularity of the matrices
\( \left( {\cal A}_{\pm }\right)  \) and \( \left( {\cal C}_{\pm }\right)  \)
is that
\begin{equation}
\label{k-descrive-stati-I}
\Pi ^{r}_{\pm }\Pi _{\pm r}=-\Pi ^{u}_{\pm }\Pi _{\pm u}-\Pi ^{i}_{\pm }\Pi _{\pm i}\neq 0
\end{equation}
where the equality is a consequence of the Virasoro constraints eq. (\ref{Virasoro-const}).
One could wonder whether this is a sufficient condition for the solvability
of eq.s (\ref{lin-comb-eps-I}), the answer is yes since would the sum of the
images of the two matrices \( \left( {\cal A}_{\pm }\right)  \) and \( \left( {\cal C}_{\pm }\right)  \)
be the whole space spanned by \( \Sigma  \) then it should be possible to find
a number \( \alpha  \) such that \( \left( {\cal A}_{\pm }+\alpha {\cal C}_{\pm }\right)  \)
is always invertible but it is easy to verify that the determinant of the previous
combination is proportional to eq. (\ref{k-descrive-stati-I}) too.

We conclude therefore that states which do not satisfy eq. (\ref{k-descrive-stati-I})
cannot be described in this gauge.

A similar argument for type II projectors gives as in (\cite{MeAds5})
\begin{equation}
\label{k-descrive-stati-II}
\Pi ^{r}_{\tau }\Pi _{\tau r}+\Pi ^{u}_{\sigma }\Pi _{\sigma u}+\Pi ^{i}_{\sigma }\Pi _{i\sigma }\neq 0
\end{equation}

\section{The lightcone GS action for the type IIb superstring on \protect\( AdS_{5}\times S_{5}\protect \).}

We want now to take the procjetor on the fermionic coordinate to be discarded
to be
\[
P=\frac{1}{\sqrt{2}}\gamma ^{0}\gamma ^{+}=\frac{1+\gamma ^{0}\gamma ^{3}}{2}\]
 in such a way that we parametrize the \( \kappa  \) gauge fixed (super)coset
representative as
\[
L=\exp \left( \begin{array}{cc}
 & \theta _{N}\\
\overline{\theta }^{N} & 
\end{array}\right) \exp \left( \begin{array}{cc}
i\frac{1+i\gamma ^{4}}{2}\gamma _{p}x^{p} & \\
 & 0_{4}
\end{array}\right) \exp \left( \begin{array}{cc}
i\frac{\gamma ^{4}}{2}\rho  & \\
 & 0_{4}
\end{array}\right) \]
where we have defined 
\[
\theta _{N}=Q\Xi _{N}=\left( \begin{array}{c}
\theta _{1N}\\
0\\
0\\
\theta _{2N}
\end{array}\right) \]
and \( \rho ,x^{p} \) (\( p=0\ldots 3 \)) are the \( AdS_{5} \) horospherical
coordinates. 

Starting from the usual formula for the coset geometry
\[
L^{-1}dL=\left( \begin{array}{cc}
i\frac{1+i\gamma ^{4}}{2}\gamma _{p}E^{p} & \Psi _{N}\\
\overline{\Psi }^{N} & 0_{4}
\end{array}\right) +spin-connection\]
a tedious computation gives the \( AdS_{5/4} \) potentials as
\begin{eqnarray*}
E^{4} & = & d\rho \\
E^{v} & = & e^{\rho }dx^{v}\: \: \: \: v=1,2,+\\
E^{-} & = & e^{\rho }\left( dx^{-}+\frac{i}{2}\overline{\theta }\gamma ^{-}d\theta \right) \\
\Psi _{N} & = & e^{-\frac{i}{2}\gamma ^{4}\rho }{\cal M}^{-1}d\theta _{N}\: \: \: {\cal M}=1+i\frac{1+i\gamma ^{4}}{2}\gamma _{p}x^{p}
\end{eqnarray*}
 in such a way that the fields entering the string action (\ref{S-primo-ordine})
can be written with the general formula of (\cite{MeAds5}) in an explicit way
as
\begin{eqnarray*}
E^{-} & = & e^{\rho }\left( dx^{-}+\frac{i}{\sqrt{2}}\theta ^{*Nn}d\theta _{Nn}\right) \\
A_{+} & = & -i\overline{\Psi }_{cN}\Psi _{M}\overline{\eta }_{c}^{N}\eta ^{M}=2e^{\rho }x^{+}d\theta ^{1}_{[N}d\theta ^{2}_{M]}\overline{\eta }_{c}^{N}\eta ^{M}\\
A_{-} & = & -A^{*}_{+}
\end{eqnarray*}
where \( \eta ^{N} \) (\( N=1\ldots 4 \)) are the 4 \( S_{5} \) c-number
killing spinors (\cite{rome})
\begin{eqnarray}
\eta ^{N} & = & \frac{1}{\sqrt{1-z^{2}}}\left( 1-z^{i}\tau _{i}\right) \epsilon ^{N}\nonumber 
\end{eqnarray}
with \( \epsilon ^{N} \) are constant 5D-spinors which satisfy the normalizarion
condition \( \epsilon ^{\dagger }_{N}\epsilon ^{M}=\delta ^{M}_{N} \) and \( \tau ^{i} \)
are the corresponding 5D \( \gamma  \) matrices and \( z^{i} \) are the \( S^{5} \)
projective coordinates .

In this way we can write the action in second order formalism as
\begin{eqnarray}
S=\frac{T}{2}\int _{\Sigma _{2}}d^{2}\xi \, \sqrt{-g}\, g^{\alpha \beta }\frac{1}{2}\left\{ \partial _{\alpha }\rho \: \partial _{\beta }\rho \right. -e^{2\rho }\: \left[ \partial _{\alpha }x^{1}\partial _{\beta }x^{1}+\partial _{\alpha }x^{2}\partial _{\beta }x^{2}\right]  &  & \nonumber \\
+2e^{2\rho }\partial _{\alpha }x^{+}\left[ \partial _{\beta }x^{-}+\frac{i}{\sqrt{2}}\theta ^{*Nm}\dd _{\beta }\theta _{Nm}\right] -\left. 4\delta _{ij}\frac{\partial _{\alpha }z^{i}\: \partial _{\beta }z^{j}}{(1-z^{2})^{2}}\right\}  &  & \nonumber \\
+\frac{e^{i\phi }}{2}e^{\rho }x^{+}d\theta ^{1}_{[N}d\theta _{M]}^{2}\overline{\eta }^{N}_{c}\eta ^{M}-\frac{e^{-i\phi }}{2}e^{\rho }x^{+}d\theta _{1}^{*[N}d\theta _{2}^{*M]}\overline{\eta }_{N}\eta _{cM} &  & \label{S-LC-horo+proj} 
\end{eqnarray}
This action can be wriite in a more compact way if we introduce the 10D 32 components
Majorana spinor defined by
\begin{equation}
\label{new-10D-spinor}
\Theta =e^{-i\frac{\pi }{4}}\left( \begin{array}{c}
e^{i\frac{\phi }{2}}\frac{1+i\gamma _{4}}{2}\theta _{N}\otimes \epsilon ^{N}+e^{-i\frac{\phi }{2}}\frac{1-i\gamma _{4}}{2}\theta _{c}^{N}\otimes \epsilon _{cN}\\
-e^{-i\frac{\phi }{2}}\frac{1-i\gamma _{4}}{2}\theta _{N}\otimes \epsilon ^{N}+e^{i\frac{\phi }{2}}\frac{1+i\gamma _{4}}{2}\theta _{c}^{N}\otimes \epsilon _{cN}
\end{array}\right) 
\end{equation}
and the coordinates 
\begin{eqnarray}
y^{4} & = & e^{\rho }\frac{1+z^{2}}{1-z^{2}}\nonumber \\
y^{i} & = & e^{\rho }\: \frac{2z^{i}}{1-z^{2}}\label{bos-chan} 
\end{eqnarray}
so that we get
\begin{eqnarray}
 & S=\frac{T}{2}\int _{\Sigma _{2}}d^{2}\xi \, \sqrt{-g}\, g^{\alpha \beta }\frac{1}{2}\left\{ \frac{\partial _{\alpha }y^{t}\: \partial _{\beta }y_{t}}{y^{2}}\right.  & \nonumber \\
 & -y^{2}\: \left[ \partial _{\alpha }x^{1}\partial _{\beta }x^{1}+\partial _{\alpha }x^{2}\partial _{\beta }x^{2}\right] +\left. 2y^{2}\partial _{\alpha }x^{+}\left[ \partial _{\beta }x^{-}+\frac{i}{4}\overline{\Theta }\Gamma ^{-}\partial _{\beta }\Theta \right] \right\}  & \nonumber \\
 & +x^{+}y^{t}d\overline{\Theta }\Gamma _{+}\Gamma _{t}d\Theta  & \nonumber \label{S-LC-x+y} 
\end{eqnarray}
 where \( t=4,5,\ldots 9 \)is the transverse coordinates label as in (\cite{kallosh-action}).

This is not the end of the story since after the \( \kappa  \) symmetry gauge
fixing there should be 16 real fermionic d.o.f. while \( \Theta  \) has 32
real components: it is infact easy to verify that the spinor \( \Theta  \)
(\ref{new-10D-spinor}) satisfies \( \frac{1}{2}\Gamma ^{-}\Gamma ^{+}\Theta =\Theta  \).
We can therefore change the \( \Gamma  \) matrices representation to the usual
lightcone basis in such a way that we can write the explicit form for the previous
projector \( \frac{1}{2}\Gamma ^{-}\Gamma ^{+}=\left( \begin{array}{cc}
1_{16} & \\
 & 0_{16}
\end{array}\right)  \) and then we can introduce a 8D spinor \( \zeta  \) so that \( \Theta =\left( \begin{array}{c}
\zeta \\
0_{16}
\end{array}\right)  \)

It is finally possible to take the flat limit rescaling \( x^{p}\rightarrow ex^{p},\rho \rightarrow e\rho ,z^{i}\rightarrow ez^{i},\Theta \rightarrow \sqrt{e}\Theta  \)
and letting \( e\rightarrow 0 \); we find that the previous action nicely reduces
to the usual 10D flat lightcone action.

\section{Conclusions.}

In this letter we have classified the possible costant projector for the fixing
of \( \kappa  \) symmetry according to their behavior under Dirac conjugation
and under charge conjugation and we have discussed the role of the different
classes. It turns out that there is a preferred class, the S class (\ref{S}),
which always yields (super)vielbeins quadratic in the fermionic coordinates.

On the other hand the behaviour under charge conjugation (\ref{I},\ref{II})
is connected to the costraint the physical states which can be described by
the projector have to satisfy (\ref{k-descrive-stati-I},\ref{k-descrive-stati-II}).

Using this knowledege we have then chosen the ``simplest'' gauge\footnote{
As a matter of facts there is another gauge fixing with the same simplicity
and it is given by \( P=\frac{1+\gamma ^{0}\gamma ^{4}}{2} \)
}, the one which corresponds to the lightcone in the flat limit, and we have
constructed the GS action for the type IIB superstring on \( AdS_{5}\times S_{5} \)
in this gauge. It turns nicely out that this action reduces to the usual GS
action for the type IIB superstring in the flat limit.While the equations of
motion cannot be solved exactly there is nevertheless hope that we can make
use of this to start an effective large radius expansion as proposed in (\cite{Rozali}).

\appendix

\section{\protect\( AdS_{5}\otimes S_{5}\protect \) conventions.\label{app_conv}}

\begin{itemize}
\item Flat indices: \( \widehat{a},\widehat{b},\ldots \in \{0,\ldots ,9\} \); \( a,b,\ldots \in \{0,1,2,3,4\} \),
\( i,j,\ldots \in \{5,\ldots ,9\} \); \( p,q,\ldots \in \{0,1,2,3\} \), \( t,u,v,\ldots \in \{4,\ldots ,9\} \);
\( \underline{\alpha },\underline{\beta },\ldots \in \left\{ 0,1\right\}  \) 
\item Flat metrics: \( \eta _{\widehat{a}\widehat{b}}=diag(+1,-1,\ldots -1) \), \( \eta _{ab}=diag(+1,-1,-1,-1,-1) \)
, \( \eta _{ij}=diag(-1,-1,-1,-1,-1) \); \( \eta _{\alpha \beta }=diag(+1,-1) \)
\item Epsilons: \( \epsilon _{0\ldots 9}=\epsilon _{0\ldots 4}=\epsilon _{5\ldots 9}=1 \);
\( \epsilon _{01}=1 \) 
\item Charge conjugate spinors: \( \Theta _{c}=C\overline{\Theta }^{T} \)
\item 1+9D gamma matrices
\begin{eqnarray*}
\{\Gamma _{\widehat{a}},\Gamma _{\widehat{b}}\} & = & 2\eta _{\widehat{a}\widehat{b}}\\
\Gamma _{11} & = & \Gamma _{0}\ldots \Gamma _{9}=\left( \begin{array}{cc}
1_{16} & \\
 & -1_{16}
\end{array}\right) \\
\Gamma _{\widehat{a}}^{T}=-\widehat{C}^{-1}\: \Gamma _{\widehat{a}}\: \widehat{C}\: \: \:  & \widehat{C}^{T}=-\widehat{C}\: \: \:  & \widehat{C}^{\dagger }=\widehat{C}^{-1}
\end{eqnarray*}
 
\item 1+4D (\( AdS_{5} \)) gamma matrices 
\begin{eqnarray*}
\{\gamma _{a},\gamma _{b}\} & = & 2\eta _{ab}\\
\gamma _{0}\gamma _{1}\gamma _{2}\gamma _{3}\gamma _{4} & = & 1_{4}\\
\gamma _{a}^{T}=+C^{-1}\: \gamma _{a}\: C\: \: \:  & C^{T}=-C\: \: \:  & C^{\dagger }=C^{-1}
\end{eqnarray*}
Moreover we use the following (1+4)D \( \gamma  \) explicit representation
\end{itemize}

\[
\gamma _{p}=\left( \begin{array}{cc}
 & \sigma _{p}\\
\widetilde{\sigma _{p}} & 
\end{array}\right) \; \gamma _{4}=\left( \begin{array}{cc}
i\, 1_{2} & \\
 & -i\, 1_{2}
\end{array}\right) \; C=\left( \begin{array}{cc}
i\sigma _{2} & \\
 & i\sigma _{2}
\end{array}\right) \]

with \( p,q,\ldots =0\ldots 3 \), \( \sigma _{p}=\left\{ 1_{2},-\sigma _{1},-\sigma _{2},-\sigma _{3}\right\}  \)
and \( \widetilde{\sigma }_{p}=\left\{ 1_{2},\sigma _{1},\sigma _{2},\sigma _{3}\right\}  \)

\begin{itemize}
\item 0+5D (\( S_{5} \)) gamma matrices
\begin{eqnarray*}
\{\tau _{i},\tau _{j}\} & = & 2\eta _{ij}\\
\tau _{5}\tau _{6}\tau _{7}\tau _{8}\tau _{9} & = & i\: 1_{2}\\
\tau _{i}^{T}=+C^{-1}\: \tau _{i}\: C & \: \: \:  & C_{5}=C\: \: \: 
\end{eqnarray*}

\item 1+9D gamma expressed through \( AdS_{5} \) \( \gamma ^{a} \) and \( S_{5} \)
\( \tau ^{i} \) gammas
\begin{eqnarray*}
\Gamma \widehat{^{a}} & = & \left\{ \gamma ^{a}\otimes 1_{4}\otimes \sigma _{1}\, ,\, 1_{4}\otimes \tau ^{i}\otimes (-\sigma _{2})\right\} \\
\widehat{C} & = & C\otimes C_{5}\otimes \sigma _{2}
\end{eqnarray*}
Moreover we use the following (1+4)D \( \gamma  \) explicit representation
\end{itemize}

\[
\gamma _{p}=\left( \begin{array}{cc}
 & \sigma _{p}\\
\widetilde{\sigma _{p}} & 
\end{array}\right) \; \gamma _{4}=\left( \begin{array}{cc}
i\, 1_{2} & \\
 & -i\, 1_{2}
\end{array}\right) \; C=\left( \begin{array}{cc}
i\sigma _{2} & \\
 & i\sigma _{2}
\end{array}\right) \]

with \( p,q,\ldots =0\ldots 3 \), \( \sigma _{p}=\left\{ 1_{2},-\sigma _{1},-\sigma _{2},-\sigma _{3}\right\}  \)
and \( \widetilde{\sigma }_{p}=\left\{ 1_{2},\sigma _{1},\sigma _{2},\sigma _{3}\right\}  \)

\begin{itemize}
\item 0+5D (\( S_{5} \)) gamma matrices
\begin{eqnarray*}
\{\tau _{i},\tau _{j}\} & = & 2\eta _{ij}\\
\tau _{5}\tau _{6}\tau _{7}\tau _{8}\tau _{9} & = & i\: 1_{2}\\
\tau _{i}^{T}=+C^{-1}\: \tau _{i}\: C & \: \: \:  & C_{5}=C\: \: \: 
\end{eqnarray*}

\item 1+9D gamma expressed through \( AdS_{5} \) \( \gamma ^{a} \) and \( S_{5} \)
\( \tau ^{i} \) gammas
\begin{eqnarray*}
\Gamma \widehat{^{a}} & = & \left\{ \gamma ^{a}\otimes 1_{4}\otimes \sigma _{1}\, ,\, 1_{4}\otimes \tau ^{i}\otimes (-\sigma _{2})\right\} \\
\widehat{C} & = & C\otimes C_{5}\otimes \sigma _{2}
\end{eqnarray*}

\end{itemize}

\section{Spinorial currents and useful Fierz identities.\label{AdS5_conv}}

We define the following useful quantities
\begin{eqnarray}
J^{i}|_{N}^{M} & = & \overline{\eta }_{N}\tau ^{i}\eta ^{M}=\overline{\eta }_{c}^{M}\tau ^{i}\eta _{cN}\label{j-inizio} \\
J^{i}|^{MN} & = & \overline{\eta }_{c}^{N}\tau ^{i}\eta ^{M}=-J^{i}|^{NM}\\
J^{i}|_{MN} & = & \overline{\eta }_{M}\tau ^{i}\eta _{cN}=-J^{i}|_{NM}\\
L^{MN} & = & \overline{\eta }_{c}^{M}\eta ^{N}=-L^{NM}\\
L_{MN} & = & \overline{\eta }_{M}\eta _{cN}=-L_{NM}\label{l-fine} 
\end{eqnarray}
where \( \eta ^{N} \) are the 4 \( S_{5} \) c-number killing spinors (\( N=1\ldots 4 \))
defined by

\[
D_{SO(5)}\eta ^{N}\equiv \left( d-\frac{1}{4}\varpi ^{ij}\tau _{ij}\right) \eta ^{N}=-\frac{e}{2}\tau _{i}\eta ^{N}E^{i}\]
 In proving that the action (\ref{S-primo-ordine}) is invariant under the \( \kappa  \)
symmetry variations given by eq.s (\ref{kappa-symmetry}) it is important to
use the following Fierz identities:
\begin{eqnarray}
J^{i}|_{N}^{M}\: L^{NP} & = & L^{MP}\: J^{i}|_{P}^{N}=J^{i}|^{MN}\label{fierz-1} \\
L^{MN}\: L_{NP} & = & \delta ^{M}_{P}\\
J^{i}|_{N}^{M}\: J^{j}|_{P}^{N}+J^{j}|_{N}^{M}\: J^{i}|_{P}^{N} & = & 2\eta ^{ij\: }\: \delta ^{M}_{P}\label{fierz-3} 
\end{eqnarray}
The derivation of these Fierz identities is based on the fact that \( \eta ^{N} \)
are a basis of the vector space on which the spinor irrep acts, i.e. \( \overline{\eta }_{N}\tau \: \eta ^{N}\propto tr(\tau ) \)
.

\end{document}